\documentclass[conference]{IEEEtran}
\IEEEoverridecommandlockouts
\usepackage{cite}
\usepackage{amsmath,amssymb,amsfonts}
\usepackage{algorithmic}
\usepackage{graphicx}
\usepackage{textcomp}
\usepackage{xcolor}
\def\BibTeX{{\rm B\kern-.05em{\sc i\kern-.025em b}\kern-.08em
    T\kern-.1667em\lower.7ex\hbox{E}\kern-.125emX}}

\usepackage{xcolor}
\usepackage[nopostdot,toc,nogroupskip,noredefwarn]{glossaries}

\newacronym{eol}{EoL}{End-of-Life}
\newacronym{sota}{SotA}{State-of-the-Art}

\newacronym{sdn}{SDN}{Software-Defined Networking}
\newacronym{qot}{QoT}{Quality of Transmission}
\newacronym{otn}{OTN}{Optical Transport Network}
\newacronym{bvt}{BVT}{Bandwidth Variable Transceiver}
\newacronym{ols}{OLS}{Optical Line System}
\newacronym{roadm}{ROADM}{Reconfigurable Optical Add-Drop Multiplexer}
\newacronym{osaas}{OSaaS}{Optical Spectrum-as-a-Service}
\newacronym{pce}{PCE}{Path Computation Element}
\newacronym{ila}{ILA}{In-line Amplifier}
\newacronym{edfa}{EDFA}{Erbium-Doped Fiber Amplifier}
\newacronym{rcsa}{RCSA}{Routing, Configuration and Spectrum Assignment}
\newacronym{art}{ART}{Aggregate Requested Traffic}
\newacronym{nlp}{NLP}{Number of in-operation Lightpaths}
\newacronym{sws}{SWS}{Single-Wavelength Source}
\newacronym{mws}{MWS}{Multi-Wavelength Source}
\newacronym{mux}{MUX}{Multiplexer}
\newacronym{demux}{DEMUX}{De-Multiplexer}
\newacronym{ba}{BA}{Booster Amplifier}
\newacronym{ca}{CA}{Comb Amplifier}
\newacronym{lp}{LP}{lightpath}

\newacronym{int}{INT}{Interfering channel}
\newacronym{cut}{CUT}{Channel-Under-Test}
\newacronym{snr}{SNR}{Signal-to-Noise Ratio}
\newacronym{gsnr}{GSNR}{Generalized Signal-to-Noise Ratio}
\newacronym{nli}{NLI}{Non-Linear Interference}
\newacronym{sci}{SCI}{Self-Channel Interference}
\newacronym{xci}{XCI}{Cross-Channel Interference}
\newacronym{mci}{MCI}{Multi-Channel Interference}
\newacronym{ase}{ASE}{Amplified Spontaneous Emissions}
\newacronym{isrs}{ISRS}{Inter-Stimulated Raman Scattering}
\newacronym{ssmf}{SSMF}{Standard Single-Mode Fiber}
\newacronym{egn}{EGN}{Enhanced Gaussian Noise}
\newacronym{gn}{GN}{Gaussian Noise}
\newacronym{cfm}{CFM}{Closed-Form Model}
\newacronym{qpsk}{QPSK}{Quadrature Phase-Shift Keying}
\newacronym{qam}{QAM}{Quadrature Amplitude Modulation}
\newacronym{ber}{BER}{Bit-Error Rate}
\newacronym{psd}{PSD}{Power Spectral Density}
\newacronym{wdm}{WDM}{Wavelength-Division Multiplexing}
\newacronym{fsr}{FSR}{Free-Spectral Range}
\newacronym{txosnr}{OSNR\textsubscript{Tx}}{transmit Optical Signal-to-Noise Ratio}
\newacronym{ocnr}{OCNR}{Optical Carrier-to-Noise Ratio}
\newacronym{iq}{I/Q}{In-phase/Quadrature}
\newacronym{sr}{SR}{symbol rate}
\newacronym{fec}{FEC}{forward-error correction}
\newacronym{ccotdr}{CC-OTDR}{coherent correlation OTDR}
\newacronym{phiotdr}{$\phi$-OTDR}{coherent optical time domain reflectometry}
\newacronym{otdr}{OTDR}{optical time domain reflectometry}
\newacronym{prbs}{PRBS}{pseudo-random binary sequence}
\newacronym{das}{DAS}{distributed acoustic sensing}
\newacronym{awg}{AWG}{arbitrary waveform generator}

\newacronym{ml}{ML}{Machine Learning}
\newacronym{xgb}{XGB}{XGBoost}
\newacronym{nn}{NN}{Neural Network}
\newacronym{ann}{ANN}{Artificial Neural Network}
\newacronym{snn}{SNN}{Self-Normalizing Neural Network}
\newacronym{rmse}{RMSE}{Root-Mean-Squared Error}
\newacronym{wm}{WM}{weighted mean}
\newacronym{relu}{ReLU}{Rectified Linear Unit}
\newacronym{selu}{SeLU}{Scaled exponential Linear Unit}
\newacronym{cdf}{CDF}{Cumulative Distribution Function}
\newacronym{mae}{MAE}{mean absolute error}

\begin{document}

\title{Low-Complexity Event Detection and Identification in Coherent Correlation OTDR Measurements\\
\thanks{This work was supported by the ECO-eNET project with funding from the Smart Networks and Services Joint
Undertaking (SNS JU) under grant agreement No. 10113933. The JU receives support from the European Union’s Horizon Europe research and innovation program. The work has been partially funded by the Horizon Europe Framework Programme under SoFiN Project (Grant No 101093015) and Villum Foundation VI-POPCOM project (VIL54486).
% German Federal Ministry of Education and Research in the project FRONT-RUNNER (Grant-IDs: \#16KISR006, \#16KISR009),
}
}

\author{\IEEEauthorblockN{Jasper M\"uller\IEEEauthorrefmark{1}\IEEEauthorrefmark{2}, Ognjen Jovanovic\IEEEauthorrefmark{1}, Florian Azendorf\IEEEauthorrefmark{1}, André Sandmann\IEEEauthorrefmark{1}, Roman Ermakov\IEEEauthorrefmark{3},\\ Sai Kireet Patri\IEEEauthorrefmark{1}, J\"org-Peter Elbers\IEEEauthorrefmark{1}, Jim Zou\IEEEauthorrefmark{1}, Darko Zibar\IEEEauthorrefmark{3} and Carmen Mas-Machuca\IEEEauthorrefmark{2}\IEEEauthorrefmark{4}} \\
\IEEEauthorblockA{\IEEEauthorrefmark{1} Adtran, Martinsried/Munich, Germany \\
% Email: \{jasper.mueller, ognjen.jovanovic, florian.azendorf, spatri, jelbers\}@adtran.com}
Email: jasper.mueller@adtran.com}
\IEEEauthorblockA{\IEEEauthorrefmark{2} Chair of Communication Networks, Technical University of Munich (TUM), Germany \\}
% Email: cmas@tum.de}
\IEEEauthorblockA{\IEEEauthorrefmark{3} Department of Electrical and Photonics Engineering, Technical University of Denmark\\}
% Email: tobias.fehenberger@advasecurity.com}}
\IEEEauthorblockA{\IEEEauthorrefmark{4} Chair of Communication Networks, Universität der Bundeswehr München, Germany\\}
% Email: tobias.fehenberger@advasecurity.com}}
}

\maketitle

\begin{abstract}
 Pairing coherent correlation OTDR with low-complexity analysis methods, we investigate the detection of fast temperature changes and vibrations in optical fibers. A localization accuracy of $\sim$2~m and extraction of vibration amplitudes and frequencies is demonstrated.%
\end{abstract}


\section{Introduction}
% Sensing Motivation
In recent years, distributed optical fiber sensing has attracted significant interest in industrial and research applications. Utilizing the inherent advantages of optical fibers, such as long reach and resilience to extreme environmental conditions, distributed sensing systems encompass a wide range of applications. 

% Importance of automated detection... algorithms
Distributed sensing systems such as \gls*{phiotdr} \cite{Lu2010jlt} and \gls*{ccotdr} \cite{sandmannOFC2023} produce measurements that contain large amounts of data and are difficult to analyze and interpret. Therefore, many applications require automated event detection and identification methods. 
% Lit research
Detection and identification of phase change events in \gls*{phiotdr} and \gls*{ccotdr} measurements are complex and challenging to generalize. Therefore, the application of \gls*{ml} algorithms has been proposed, demonstrating the potential of deep learning \cite{aktas2017deep, shiloh2018deep} and, specifically, convolutional neural networks \cite{shi2019event, wu2019one} for the analysis of \gls*{phiotdr} measurements. Efforts to reduce the required amount of labeled training data have utilized transfer learning \cite{shi2021easy} or generative models to increase the dataset size \cite{shi2022event}. Feature extraction-based methods have also been proposed \cite{zhu2014vibration, jiang2018event}, but require careful feature selection and complex processing. 

% We do CC OTDR, not Phi OTDR
While previous work focused on event detection and identification in \gls*{phiotdr} systems, in this paper we use a \gls*{ccotdr} system for increased spatial resolution or reach. 
% Motivation for our low-complexity approach
The difficulty in applying \gls*{ml} techniques lies in the generation of suitable data sets. It is challenging to generate a realistic simulation of a \gls*{ccotdr} measurement due to dynamic disturbances on the fiber and its different coating layers 
\cite{SimulationModel2015}. This leaves experimental setups as the only viable option, leading to particularly time-intensive data generation.

% We don't need ML for everything
Several use-cases of distributed sensing and \gls*{ccotdr}, such as traffic monitoring, distributed temperature monitoring of, e.g., power lines, structural health monitoring, perimeter and border protection, leakage detection, etc., focus on detecting specific events defined before the system is deployed \cite{wellbrock2022ofc}. Therefore, in these cases, the generalizability of the algorithms can be traded off with lower requirements in training dataset size and algorithm complexity.

% In this work...
In this work, we present low-complexity methods for event detection and identification. The presented methods do not require large datasets for training but only a few data points for parameter tuning. We evaluate the detection and identification methods on an experimental data set collected through multiple measurement series. We reliably detect temperature changes and vibrations in the experimental setup. A high localization accuracy of temperature change events with a standard deviation of 2.7~m is achieved. A simple approach to fit a meaningful function to the observed phase changes enables the discrimination of temperature change and vibration events in the \gls*{ccotdr} measurements. 
% strain and vibration? maybe [temperature and vibration events] 
It allows for further characterization, such as determining the amplitude and frequency of vibration events, to detect whether appliances or machines in the area are active.

\begin{figure*}[htp!]
    \centering
    \includegraphics[width=0.95\linewidth]{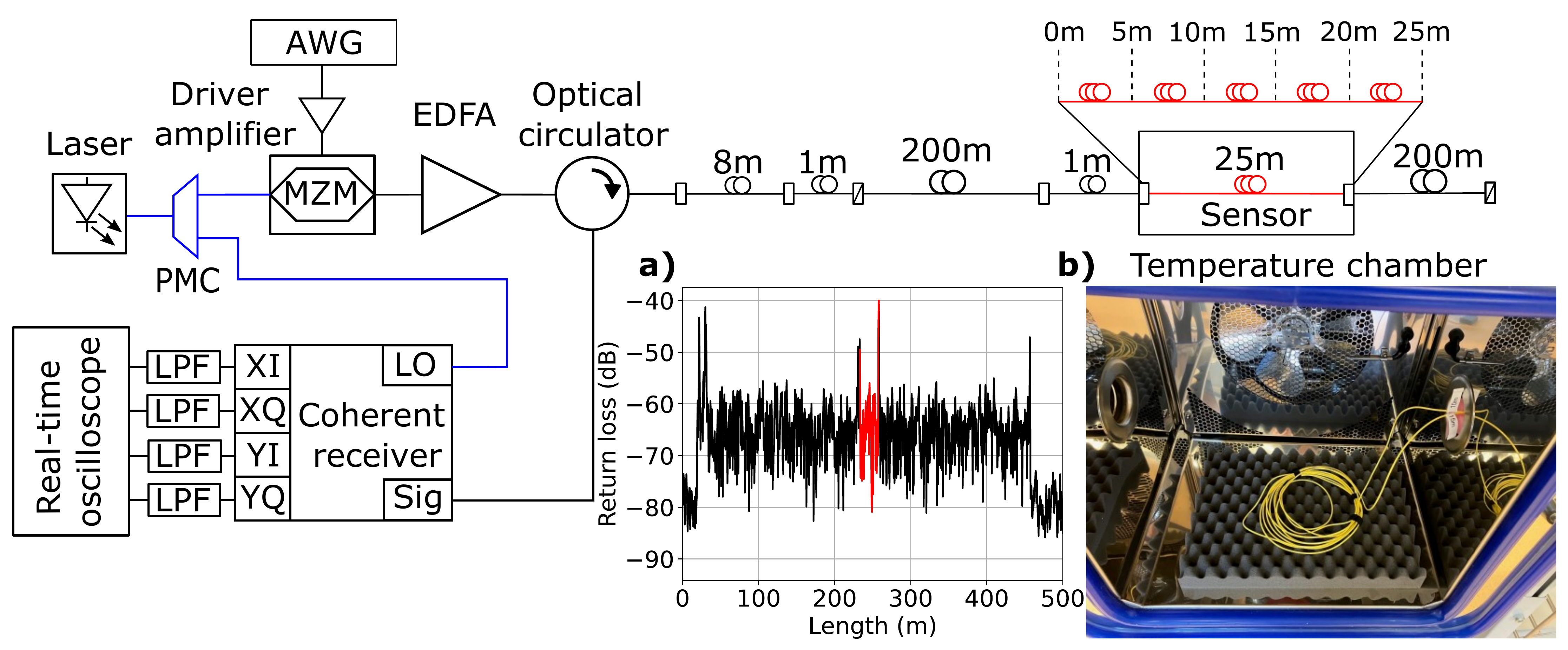}
    \caption{Coherent correlation OTDR experimental setup with the 25~m sensor fiber (colored in red) is separated into 5~m segments. \textbf{a)} Corresponding fiber fingerprint. \textbf{b)} One or more segments are placed inside the temperature chamber.}
    \label{fig:exp_setup}
\end{figure*}

\section{Experimental Setup}
Fig.~\ref{fig:exp_setup} illustrates the experimental setup of the \gls*{ccotdr}. A continuous wave signal of a highly coherent laser with a Lorentzian linewidth of less than 100 Hz is sent into a polarization-maintaining coupler, which equally splits the optical power into two output branches. Subsequently, one part is fed into the local oscillator port of the coherent receiver. In contrast, the phase of the other part is modulated with a Mach-Zehnder modulator at a bit rate of 125~Mbit/s, leading to a spatial resolution of 80~cm. The probe signal is generated by an \gls*{awg}. It consists of a 4095-bit pseudo-random binary sequence, a trailing “-1” symbol, and a zero-padding of 5000~symbols, yielding a frame duration of 72.8~µs. Afterwards, the probe signal is amplified with an Erbium-doped fiber amplifier and coupled into the standard single-mode fiber via an optical circulator. In the experiment, two 200~m of lead-in and termination fibers are interconnected with a 25~m patch cord used as a sensor. It is worth mentioning that this sensor fiber has a 3 mm jacket (yellow-colored) to protect the fiber, as shown in Fig.~\ref{fig:exp_setup}~b). The sensor fiber was placed in a temperature-controlled cabinet to simulate temperature variations in the vicinity of the fiber, as shown in the inset. The 25~m sensor (red) was divided into 5-meter segments to reduce or increase the temperature impacts along the fiber. At the end of the termination fiber, an angled physical contact connector is installed to mitigate saturations at the receiver. The backscattered and reflected signals are received at the signal input of the coherent receiver. Applying the self-homodyne reception scheme, amplitude, phase, and polarization information are extracted. A real-time oscilloscope records the four field components with a sampling rate of 625 MS/s and stores them for offline processing. These signals are cross-correlated with the transmitted sequence. Here, the advantage of the correlation approach is that the signal-to-noise ratio is improved compared with a single pulse approach in conventional $\phi$-OTDR, while the spatial resolution is maintained. Fig.~\ref{fig:exp_setup}~a) shows the optical return loss of the fiber. Reflections at connector pairs cause the peaks with amplitudes of -38 dB to -48 dB, whereas all the other peaks correspond to the Rayleigh backscattering interference pattern.

\begin{figure}[t]
    \centering
    \includegraphics[width=\linewidth]{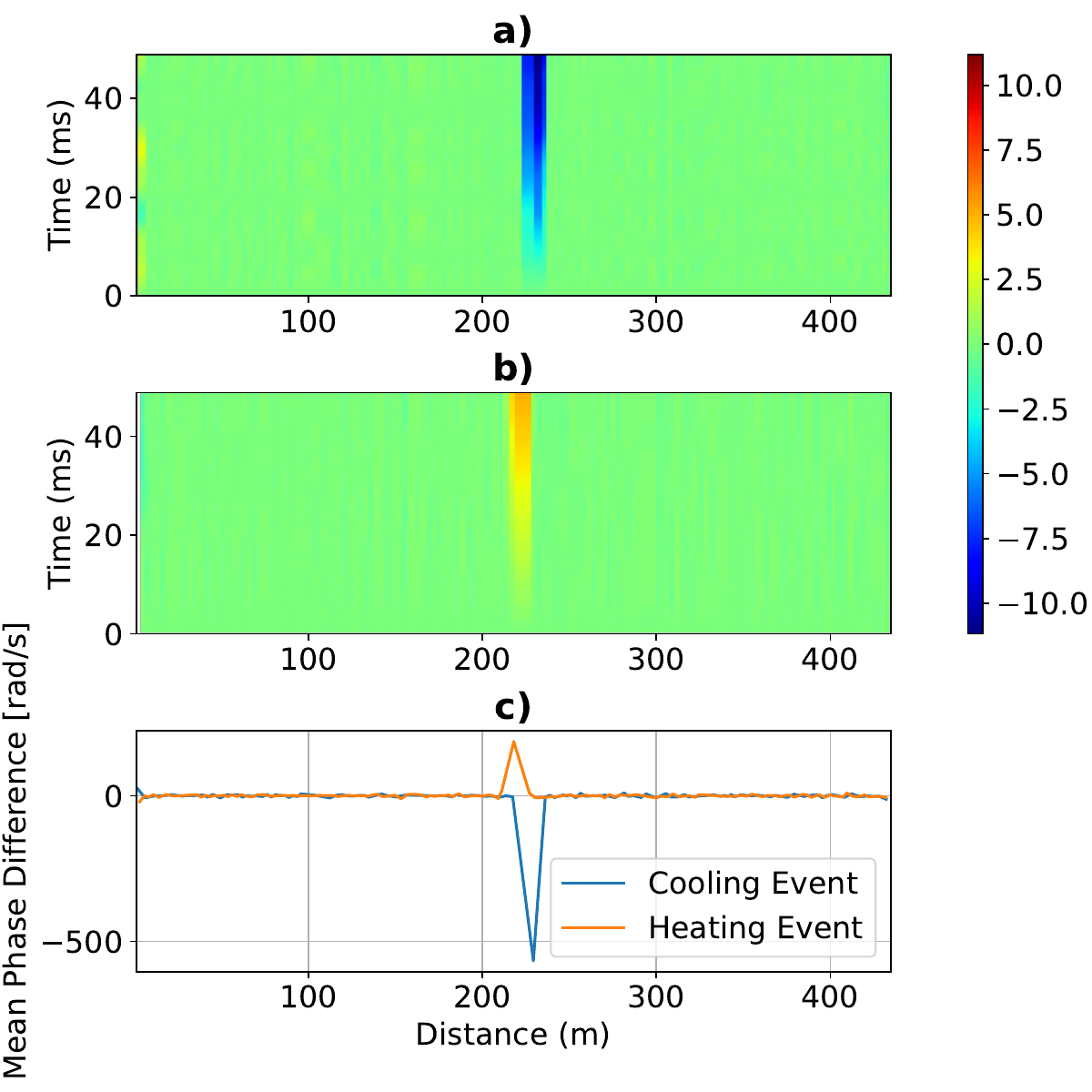}
    \caption{Phase difference waterfall of an example \textbf{a)} cooling event with a temperature change of -1.6$^\circ\text{C}$, \textbf{b)} heating event of +0.9$^\circ\text{C}$ and \textbf{c)} mean phase difference of the events.}
    \label{fig:detection}
\end{figure}

\begin{figure*}[htp!]
    \centering
    \includegraphics[width=\linewidth]{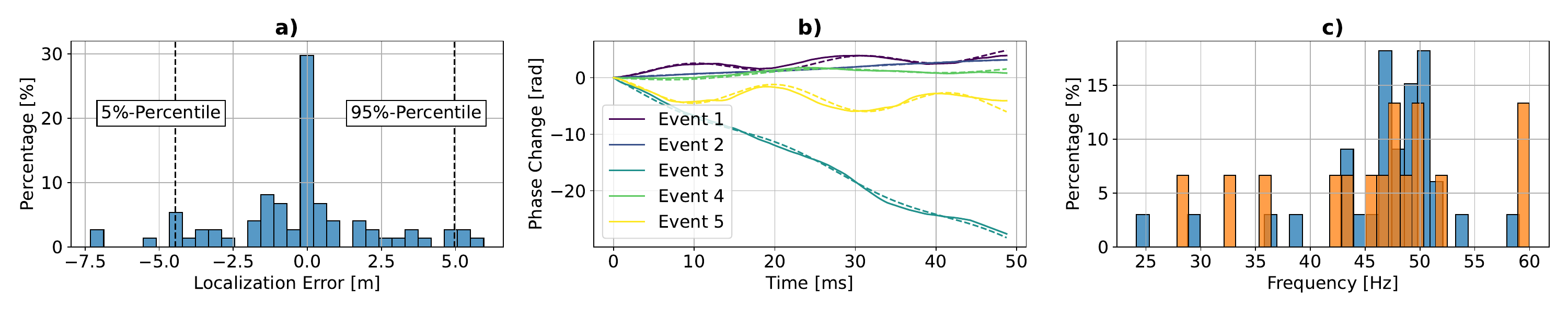}
    \caption{\textbf{a)} Temperature change event localization error distribution, \textbf{b)} phase change function fit of vibration and temperature change events, \textbf{c)} detected frequencies of vibration events in heating (orange) and cooling (blue) cycle.}
    \label{fig:results}
\end{figure*}

% Measurements
\section{Data Generation}
% What we do
We aim to utilize analytical methods that can be tuned to the experimental setup using only a few data points for calibration. For the evaluation of these techniques, measurement series were conducted by placing one or multiple 5-meter segments of the sensor fiber into the temperature chamber while heating up the chamber from a room temperature of $\sim\text{25}^\circ\text{C}$ up to $\sim\text{40}^\circ\text{C}$. During the heating period of around 250~s, regular measurements of 50~ms each have been taken with a time gap of $\sim$ 28~s between the individual measurements. The regular measurements continue after the heating period ends to capture the cooling process. In total, 125 measurements were considered for the evaluation of event detection and discrimination methods.

% Event detection and localization
\section{Event Detection}
% Resolution
The detection of phase change events, such as temperature change, or vibration, is performed by estimating the phase difference between consecutive fingerprint peaks in combination with a simple threshold. The accuracy of such event detection highly depends on the chosen threshold, the considered peaks, and their location. The peaks are selected by defining a minimum magnitude of a peak and a minimum distance between consecutive peaks. A trade-off between localization and event detection accuracy arises depending on the chosen minimum distance. As a result of a coarse optimization, a spatial resolution of 2~m was chosen, and an example of a phase waterfall plot is shown in Fig.~\ref{fig:detection}~a) for a cooling event and Fig.~\ref{fig:detection}~b) for a heating event.
In Fig.~\ref{fig:detection} c), the average of the phase increments over time is shown. On average, there is no phase change outside of the event, whereas there is a spike in the area of a temperature change event. Vibrations are instead detected through the variance of the phase increments.
For the given bit rate, this method achieves an accurate localization with a mean absolute error of 1.8~m, a standard deviation of 2.7~m, and 90~\% of the localization within 5~m of the event location as shown in Fig.~\ref{fig:results} a). 
% The localization error improves if the event starts or ends at a reflective event peak (e.g. from a fiber connector). An \gls*{mae} of 1.3~m is achieved in this case. %It was noticed that choosing a larger minimum distance between consecutive peaks would achieve a more accurate event detection but at the expense of a less accurate event localization. A higher localization accuracy would be achieved for shorter distances, but event detection would be less accurate.
%For accurate phase estimation, peaks in the fingerprint are utilized as positions for phase evaluation \jm{fix this}. Therefore, the achievable spatial resolution for phase change event detection depends on the \gls*{snr} and the distance between \gls*{snr} peaks in the fingerprint of the area under test. We target a spatial resolution of 2m in the given setup (Fig.~\ref{fig:exp_setup}).
% Event localization

% Event discrimination
\section{Event Characterization}
Once detected, we aim to extract further insights from the phase change event. Prior knowledge of the expected events for a given use case can significantly simplify event identification and reduce the complexity of the identification algorithm. While complex \gls*{ml} models are commonly used for the identification (classification) task \cite{tejedor2017machine}, we aim to showcase the effectiveness of a simple function fit for event identification in \gls*{ccotdr} measurements. 
In the presented use case, the temperature chamber is the area under test. We expect phase change events related to the temperature change and the vibrations of the cooling compressor.
We assume the restriction of a single vibration event in the measurement at most to limit the complexity of the fitted function, defined as:
\begin{equation}
    f(t) = C t + A  \sin\left(2\pi f t + \phi_0\right) - A  \sin\left(\phi_0\right) 
\end{equation}
where $C$ is the slope of the phase change, $A$ is the amplitude of the sinusoidal component and $\phi_0$ is the phase offset. The term $- A \sin\left(\phi_0\right)$ ensures $f(0) = 0$. Fig.~\ref{fig:results} b) shows examples of detected phase change events and the function fit.

% distinguishing temperature change (strain) and vibration
The fitted parameters can be used to characterize the detected events. The slope and amplitude represent the strength of strain or temperature change and vibration events, respectively, and therefore allow the categorization of events.
The frequency and amplitude of vibration events further characterize vibration events, allowing for the detection of distinct vibration sources. Fig.~\ref{fig:results}~c) shows the recorded frequencies of the vibration events in the measurement data. Due to acoustic noise in the laboratory and only segments of the sensor fiber being isolated in the temperature chamber, a range of frequencies is measured. While the distribution of frequencies in the heating phase is fairly even between the occurring vibration frequencies, in the cooling period, most recorded frequencies are concentrated around 48~Hz, as the cooling compressor vibrates at this frequency. Some measurements in the cooling period still pick up other frequencies due to ambient noise in the laboratory after the compressor turns off. 

The slope parameter characterizes the strength of temperature events and can be used to estimate occurring temperature changes. The temperature evolution observed by phase change in the fiber core is delayed compared to the temperature sensor measurements. The delay can be described by a filter function~\cite{sandmannOFC2023}. The available data does not allow for a robust filter fit as the measurements are short (50~ms) and far apart (28~s). However, dependable detection of events with large temperature changes of more than $\text{0.1}^\circ\text{C/s}$ is achieved with 95~\% accuracy. The falsely classified cases are explained by the lag between the temperature observed at the sensor and at the fiber core.

% \section{Temperature Estimation}
% Temperature changes in the fiber can be computed from phase changes~\cite{sandmannOFC2023}, arriving at a phase temperature coefficient of 42 rad/mK for the given experimental setup. The temperature changes observed in the fiber core through phase changes are delayed compared to the temperature sensor measurements. The relation can be described by a first-order low-pass filter as shown in Fig.~\ref{fig:results} c).  

\section{Conclusions}
We achieve robust event detection and identification by utilizing low-complexity data analysis methods, thereby minimizing the reliance on extensive datasets for machine learning model training. Our approach demonstrates reliable detection of temperature change and vibration events on a data set generated by an experimental setup. We achieve a high localization accuracy for temperature change events, with a standard deviation of 2.7~m. Furthermore, our presented method differentiates temperature and vibration events in distributed optical fiber sensing measurements. Additionally, it enables the quantification of temperature events and characterization of the amplitude and frequency of vibration events. We highlight the potential of low-complexity event detection and identification methods for \gls*{ccotdr} measurement, allowing for real-time monitoring and analysis in diverse applications of distributed optical fiber sensing.

% \section*{Acknowledgment}

% The preferred spelling of the word ``acknowledgment'' in America is without 
% an ``e'' after the ``g''. Avoid the stilted expression ``one of us (R. B. 
% G.) thanks $\ldots$''. Instead, try ``R. B. G. thanks$\ldots$''. Put sponsor 
% acknowledgments in the unnumbered footnote on the first page.

\bibliographystyle{ieeetr}
\bibliography{references.bib}

\end{document}